\documentclass[preprint]{revtex4}
\usepackage{amsfonts}
\usepackage{amssymb}
\usepackage{amsmath}
\usepackage{graphicx}
\usepackage{array}
\usepackage{dcolumn}
\usepackage{color}
\newcommand{\WAC}[1]{$(H_2O)_{#1}$}

\begin{document}
\title{Monomer Basis Representation Method For Calculating The Spectra Of 
         Molecular Clusters I. The Method And Qualitative Models.}
\author{Mahir E. Ocak}
\email{meocak@alumni.uchicago.edu}
\affiliation{Ya\c{s}amkent Mahallesi, Yonca Sitesi 13/B Daire No:5, \c{C}ayyolu,
		Ankara, Turkey
            }
\date{\today}
\begin{abstract}
Firstly, a sequential symmetry adaptation procedure is derived for 
semidirect product groups. Then, this sequential symmetry adaptation 
procedure is used in the development of 
 new method named Monomer Basis Representation (MBR)
for calculating the vibration-rotation-tunneling (VRT)
spectra  of molecular clusters.  The method is based
on generation of optimized bases for each monomer in the cluster as
a linear combination of some primitive basis functions and then
using the sequential symmetry adaptation procedure for generating
a small symmetry adapted basis for the solution of the full problem.
It is seen that given an optimized basis for each monomer the application
of the sequential symmetry adaptation procedure leads to a generalized
eigenvalue problem instead of a standard eigenvalue problem if the
procedure is used as it is. In this
paper, MBR method
 will be developed as a solution of that problem such that
 it leads to  generation
of an orthogonal optimized basis for the cluster being studied
regardless of the nature of the
primitive bases that are used in the generation of optimized bases
of the monomers.
\end{abstract}
\maketitle

\section{\label{sec:intro}Introduction}
Obtaining the bulk properties of condensed phases 
from microscopic considerations
is a long standing
dream. For that purpose, it is essential to develop an accurate potential
surface.

The total potential energy function $V(\vec{r_1},\vec{r_2},\ldots,\vec{r_N})$
of a system of $N$ identical particles
can be expanded as sum of $n$-body potentials:
\begin{eqnarray}
V(\vec{r_1},\vec{r_2},\ldots,\vec{r_N})=\sum_{i_1<i_2}^N V_2(\vec{r_{i_1}},\vec{r_{i_2}})+
\sum_{i_1<i_2<i_3}^N V_3(\vec{r_{i_1}},\vec{r_{i_2}},\vec{r_{i_3}})\nonumber \\
+ \ldots +
\sum_{i_1<i_2<i_3,\ldots,i_N}^N V_n(\vec{r_{i_1}},\vec{r_{i_2}},\vec{r_{i_3}},\ldots,\vec{r_{i_N}}).
\label{expansion}
\end{eqnarray}
It has been assumed in the past that this series converges rapidly and
in most systems the pairwise additive approximation, i.e.
\begin{equation}
V(\vec{r_1},\vec{r_2},\ldots,\vec{r_N})=\sum_{i_1<i_2}^N V_2(\vec{r_{i_1}},\vec{r_{i_2}}),
\end{equation}
 has been shown to be qualitatively valid. However, this is
not the case for many systems
\cite{saykally94,moszynisky94,wormer2000,trimerrev}.
 For example, the higher order terms in equation
 (\ref{expansion})
 may add up to 25\% of the interaction energy of bulk water
\cite{hermansson94}, most of which results from  three-body effects. It has
been found that \cite{wormer99} the pair interaction energy represents
$83-86 \% $ of the total interaction energy of  \WAC{3} $\approx 75\%$
of \WAC{4} and only $\approx 68 \%$ of \WAC{5}. Therefore, it is evident that
many-body terms are far from negligible.

Before the studies of van der Waals molecules, most experiments sensitive to
many-body effects were performed on macroscopic systems with the result
that only the {\it total} many body effects were obtained. Then, condensed
phase properties have been used to determine ``effective'' pair potentials
\cite{Sinanoglu67}
which include all of the terms in the many-body expansion in an average
manner, at least over some range of temperature.

On the other hand, studies of  small clusters proved to be ideal for studying
many-body terms, because each term in the many-body expansion can be
studied uniquely. For example, dimers are ideal for studying
two-body potentials, since there are no many-body terms in
their potential surfaces. Similarly, once the potential surface of
the dimer of a substance  is known accurately,
the trimer of that substance
becomes the model system for studying three-body terms in its potential surface,
 because
three-body terms are the only many-body terms in the potential surface of
a trimer. Thus, in principle
 it is possible to develop accurate potential surfaces
by starting from dimers and deriving the many-body terms by studying bigger
and bigger clusters. However, it is very hard to accomplish in practice.
In order to develop a potential surface or in order to test the accuracy of
an available potential surface, it is necessary to make quantum mechanical
calculations.
In quantum mechanical spectra calculations, the eigenstates of the
Hamiltonian operator are usually
obtained by diagonalizing the matrix representing
the Hamiltonian operator in a finite basis and
it is well known that the size of
the basis required for the calculations grows exponentially with the number of
degrees of freedom. As a result of that, the quantum mechanical
calculations becomes harder as the size of the problem gets bigger.

Considering inter-molecular vibrations of molecular clusters (consisting of
nonlinear monomers),
dimers are six dimensional systems and trimers are twelve dimensional systems.
If it is possible to obtain accurate results with $N$ basis functions per
degree of freedom, then a dimer calculation requires $N^6$ basis functions
and a trimer calculation requires $N^{12}$ basis functions. Even
if it would be possible to obtain good results with a
small number of basis functions per degree of freedom such as $N=10$,
the size of a
 basis required for a  trimer calculation would be a million times bigger than
the size of a
 basis required for a dimer calculation. It is this exponential scaling
of the number of the basis functions which makes
the calculations harder and harder
as the size of the problem gets bigger, and obviously it is necessary to
develop methods for reducing the size of the bases.

One way of reducing the sizes of bases is to make symmetry adaptation of
basis functions and solve for each symmetry separately. However, the
well-known method of symmetry adaptation is not very helpful. 
Although, the sizes of bases grow exponentially, orders of molecular
symmetry groups \cite{Bunker79} grow linearly. For example, order
of the molecular symmetry group of water dimer is $16$ \cite{dyke77} and the 
order of the molecular symmetry group of water trimer
 is $48$ \cite{trimerI:96}.

A much more efficient way of reducing the computational cost is to use
bases that are optimized for the particular problem at hand instead of
primitive bases. 
The inefficiency of primitive bases results from the fact that 
they do not know anything about the potential surface of the system. 
The optimized bases that know about the underlying potential 
surface can be generated as linear combinations of some
primitive basis functions by taking a model potential surface
that resembles to the actual potential surface as much as possible
and solving for its eigenstates. This obviously makes it necessary 
to divide the problem into smaller parts since trying to find an
optimized basis for the full problem is as difficult as solving it.
In the case of molecular clusters an obvious way of dividing the problem
into smaller parts is to consider each monomer separately.
In the rest of the paper, the main discussion will be about
how optimized bases for each monomer can be generated 
and how they can be combined for the solution of the full problem. 

In section \ref{sec:ssa}, a sequential symmetry adaptation procedure
will be derived. By finding the relations between the projection operators
of the irreducible representations of the molecular symmetry group of
the cluster and the projection operators of the irreducible representations
of its subgroups, symmetries of monomer basis functions will be related to the 
symmetries of eigenstates of the cluster.  
This symmetry adaptation procedure will be used in section \ref{sec:mbr}
in which generation of optimized monomer bases is discussed. It will be seen
that generation of optimized bases creates its own problems related with
symmetry adaptation. In order to guarantee generation of an efficient 
orthonormal basis, it will be necessary to modify the sequential symmetry
adaptation procedure developed in section \ref{sec:ssa}.
 In order to illustrate how the method can be used for generating
small symmetry adapted bases,
two qualitative models will be given in section \ref{sec:quamod}, one
for water dimer and one for water trimer.
The method will be applied to water dimer in a forthcoming paper in order to
illustrate its application and to show its validity.

\section{\label{sec:ssa}Sequential Symmetry Adaptation}

An analysis of the structure of the molecular symmetry groups shows that
the molecular symmetry group of a molecular cluster 
consisting of $n$ non-reacting
monomers can be written in terms of its subgroups as \cite{MEOPhd} 
\begin{equation}
G(MS)= \left( \left( G_{k_1} \otimes G_{k_2} \otimes \ldots \otimes
                  G_{k_n} \right) \circledS G_l 
                     \right) \otimes \varepsilon.
\label{eq:gms}
\end{equation}
In the equation above, the groups $G_{k_i}$ with $i=1,\ldots,n$ are the
pure permutation subgroups of the monomers that have orders $k_i$; 
the group $G_l$ which is of the order of $l$ is the subgroup containing
the operations permuting the identical monomers; and the group $\varepsilon$
is the inversion subgroup that contains the identity element $E$ and the 
inversion element $E^*$.

In equation (\ref{eq:gms}), $\otimes$ denotes a direct product multiplication
and $\circledS$ denotes a semi-direct product multiplication. The difference
between a direct product and a semi-direct product multiplication is that
in a direct product multiplication both of the subgroups are invariant
subgroups of the product group while in a semi-direct product multiplication
only one of them is an invariant subgroup of the product group. Since
the operations permuting the identical monomers bring in non-commutation,
presence of a semi-direct product multiplication is inevitable.  
In this paper, it will be assumed that the group $G_l$ is cyclic
and has the order $l=n_m$. In the case of dimers the order of the
group which contains the permutations of identical monomers
 has to be 2 anyway (since $n_m ! =n_m$ for $n_m=2$), so that
 there is no assumption. However, in the case of bigger clusters,
physical meaning of this assumption is that the cluster is assumed to
have a rigid cyclic skeleton.  The reason for this assumption
will be clear in section \ref{ssec:gmbom}. It should be noted that
equation (\ref{eq:gms}) assumes that the inversion operation is a feasible
symmetry operation of the molecular symmetry group which is not true in
general. However, this corresponds to a more general case. The method
developed in this paper is applicable to any cluster whose molecular
symmetry group does not include inversion operation, too. 

Symmetry adaptation of basis functions to an irreducible representation
$\Gamma$ of a group $G$ can be done by application of the projection
operator of that irreducible representation which is given by
\cite{bunker98,Cotton}
\begin{equation}
\hat{P}^{\Gamma}_G = \frac{d_{\Gamma}}{\vert G \vert} \sum_{g \in G} 
			\chi^{\Gamma}[g]^* \hat{O}_g,
\label{eq:projop}
\end{equation}
where $g$ are the elements of the group $G$, $d_{\Gamma}$ is the dimension
of the irreducible representation $\Gamma$, $\vert G \vert$ is the order
of the group $G$, $\hat{O}_g$ is the operator representing $g$, and 
$\chi^{\Gamma}[g]$ is the character of $g$ in the irreducible representation
$\Gamma$.
As shown in the appendix, for a group $G$
that can be written as a semi-direct product of two of its subgroups
$H$ and $K$, and satisfying 
the condition given in equation (\ref{eq:cijk}), the projection operator of 
an irreducible representation can be 
decomposed into product of two terms such that
\begin{equation}
\hat{P}^{\Gamma}_G = \left( \frac{1}{\vert H \vert} \sum_{h \in H}
			\chi^{\Gamma}[h]^*\hat{O}_h \right)
			\left( \frac{1}{\vert K \vert} 
			\sum_{k \in K} \chi^{\Gamma}[k]^* \hat{O}_k \right),
\label{eq:pod}
\end{equation}
where the definitions of the terms are similar to those of equation
(\ref{eq:projop}).
Please note that the condition given in equation (\ref{eq:cijk})
 is always satisfied
for any direct product group. Therefore, equation (\ref{eq:pod}) applies
to all of the group multiplications  in equation (\ref{eq:gms}).
Consequently, it follows that the symmetry adaptation of basis functions
can be done in $n+2$ steps sequentially. Furthermore, the
characters in equation (\ref{eq:pod}) can be decomposed 
to their irreducible components in groups $H$ and $K$. Thus, if the 
character in the first parentheses in equation (\ref{eq:pod}), can be
expressed as $a_1 \Gamma_1 \oplus a_2 \Gamma_2 \oplus \ldots$ where
$a_i$ are nonnegative 
integers and $\Gamma_i$ are the irreducible representations
of the group $H$, then from the definition of projection operators
it follows that this parentheses can be expressed in terms of the 
projection operators of the irreducible representations of the group $H$
as 
\begin{equation}
 \frac{1}{\vert H \vert} \sum_{h \in H}
			\chi^{\Gamma}[h]^*\hat{O}_h =
\frac{a_1}{d_1} \hat{P}^{\Gamma_1}_H + 
\frac{a_2}{d_2} \hat{P}^{\Gamma_2}_H + \ldots, 
\label{eq:podcor}
\end{equation}
where $\hat{P}^{\Gamma_i}_H$ is the projection operator and $d_i$ are the
dimension of the irreducible
representation $\Gamma_i$. A similar expression can also be written for the 
second parentheses in equation (\ref{eq:pod}) in terms of the projection
operators of the irreducible representations of the group $K$. 
Obviously only the terms for which $a_i$ is nonzero will contribute to the sum.
Consequently, equation (\ref{eq:podcor}), or equally finding the coefficients
$a_i$ is useful for determining which symmetries of the monomer basis functions
can be symmetry adapted to an irreducible representation of the product group.

\section{\label{sec:mbr}Monomer Basis Representation Method}

If the symmetry adaptation procedure given in previous section is used 
with primitive bases, it can not provide any optimization 
more than what one can achieve with the direct symmetry adaptation by using
equation (\ref{eq:projop}).
On the other hand, sequential 
symmetry adaptation procedure combined with the physically meaningful 
partitioning of the molecular symmetry groups given in 
equation (\ref{eq:gms}) makes it possible
to devise algorithms for obtaining symmetry adapted optimized bases. 

On the other hand, generation of optimized bases 
creates its own problems  related with symmetry adaptation. 
Primitive basis functions (plane waves,
spherical harmonics, Wigner rotation functions \ldots) are the solutions
of Hamiltonians corresponding to motions of free particles or free bodies.
Since the kinetic energy is always absolutely symmetric, a free particle
Hamiltonian has absolute symmetry, too. Consequently, the basis functions
that are obtained as solutions of that Hamiltonian also have absolute 
symmetry.  As a result of that application
of a symmetry operation to a primitive basis function always results in
another function in the same basis other than a possible phase factor.
Thus the symmetry adapted basis functions can be obtained as linear
combinations of primitive basis functions. 

The case of optimized basis functions is different. An optimized basis
function should know about the particular problem at hand so that the 
Hamiltonian, that the optimized basis functions are solutions of, should
include a potential energy function. 
Since the potential surfaces do not have
absolute symmetry, a Hamiltonian including a potential energy
function can not be absolutely symmetric, either. This will restrict the
symmetries of optimized basis functions that are obtained as solutions of 
the Hamiltonian. This means that application of a symmetry operation
to an optimized basis function will not necessarily result in another
basis function in the same optimized basis. As a result of that 
solving a problem might be impossible when optimized basis 
functions are used unless a special care is taken to ensure that
the physically meaningful 
solutions  of the Hamiltonian can be obtained as linear combinations
of the optimized basis functions.

A discussion of how such optimized monomer bases can be found and how
they can be combined for the solution of the full problem will be 
made in the following subsections. 
 The procedure for obtaining the results can be divided into four steps. 
These steps are: generation of a basis for one of the monomers, generation
of bases for other monomers, combining monomer bases to generate a basis
for the full problem, and the solution of the full problem. 

Before starting to develop the method, it should be noted that a basis
function related with a monomer is a function 
 which describes the orientation of the monomer in the cluster and 
a function related with inter-monomer coordinates is a function which 
describes the orientation of the monomers with respect to each other.
The monomers
are assumed to be rigid bodies so that intra-monomer degrees of freedom
are not considered. In the discussion, language of Permutation Inversion (PI)
group theory \cite{Bunker79,bunker98,Bunker2005} will be used since it provides
the most natural way of handling the symmetries in molecular systems.

\subsection{\label{ssec:gmb}Generation of A Monomer Basis}

In order to have a basis which is fully symmetry adapted, or can be 
fully symmetry adapted, to the  molecular symmetry group of the cluster, 
there should be a basis for each monomer, which is symmetry adapted to the 
irreducible representations of the pure permutation group of the monomer $G_k$. This can be done by just taking a 
primitive basis for each monomer, 
and then symmetry adapting it to the irreducible representations
of the permutation group, $G_k$. However, in order to have an efficient 
basis for the full problem, the size of the monomer basis should be small; 
 and just taking a small  number of primitive basis 
functions for each monomer will not be a good  choice obviously. Therefore,
instead of using a primitive basis for each monomer, which does not 
know anything about the problem at hand,  it is better to have
 a basis which is optimized for the particular problem being studied. 

An optimized basis for a monomer can be generated by taking a model 
Hamiltonian for that monomer and then solving for the eigenstates of 
the model Hamiltonian with a basis which has the required symmetry
properties. Then, a small number of the 
 eigenstates of the model Hamiltonian  can be taken as an optimized basis for
that monomer. 
The model Hamiltonian should include the kinetic energy operator related
with that monomer in the Hamiltonian of the cluster 
and a model potential surface for that  monomer. 
The model potential surface can be chosen as desired. However, the
efficiency of the resulting basis will depend 
on this choice. 

The primitive basis functions which are used to generate the optimized basis
 should be  symmetry adapted by using the operations contained in the 
pure permutation group related with that monomer. According to equation
\ref{eq:pod},
this is certainly sufficient for making sequential symmetry adaptation
properly. However, in order to guarantee invariance
of the cluster basis while combining the monomer bases, it is 
better to follow a more complex path for making sequential symmetry adaptation
of basis functions because of the reasons explained below. 

If the sequential symmetry adaptation procedure is used as it is,
then the basis functions are going to be symmetry adapted to the 
irreducible representations of the inversion subgroup while forming
the symmetry adapted cluster basis functions. At this step, it will
be necessary to apply inversion operation, $E^*$ to the monomer basis
functions. When the inversion operation is applied to a basis function
of a monomer, the result will be another basis function for the same
monomer. However, the resulting function will not be in the same 
basis unless the basis has the inversion symmetry of the {\it cluster}. 
Since the optimized basis functions of the monomers are going to be 
generated with a calculation, the basis functions cannot have such a 
symmetry unless the model potential surface of the cluster has the
inversion symmetry of the cluster.  A model potential
surface for a monomer cannot have such a property unless it is 
imposed to it. Consequently, if the sequential 
symmetry adaptation procedure is used as it is, it will be necessary 
to deal with a generalized eigenvalue problem instead of a standard
eigenvalue problem. 

This problem can be overcome by using the properties of direct product
groups. The inversion subgroup can always be multiplied with 
another subgroup of the molecular symmetry group with direct product  
multiplication. If the basis functions that are used to generate optimized
monomer bases are symmetry adapted to irreducible representations of 
the direct product group obtained
from the pure permutation group of the monomer and the inversion 
subgroup of the cluster. Then, they will be symmetry adapted to the 
irreducible representations of both the pure permutation group of the monomer
and the inversion subgroup of the cluster. In this case, the application
of the inversion operation to optimized basis functions will not 
create new functions. In fact, they will be the eigenstates of the 
inversion operation with the eigenvalues $\pm 1$. If the same thing is
true, for all of the monomer bases, then the product basis of the monomer
bases will also be eigenstates of the inversion operation with the eigenvalues
$\pm 1$. Consequently, the product basis of the monomer bases will be invariant
under the effect of the inversion operation so that the symmetry adaptation
of basis functions will not lead to a generalized eigenvalue problem but 
to a standard eigenvalue problem. 

To sum up, in order to generate a small  basis for a monomer, 
the permutation group of that monomer, $G_k$, and the inversion subgroup,
$\varepsilon$, of the molecular symmetry group of the {\it cluster}
are taken and the direct product group of these two subgroups are formed.
 Then, the
eigenstates of the model Hamiltonian is solved for each 
symmetry separately after the basis functions are symmetry adapted.
 A subset of these eigenstates 
becomes the contracted basis for that monomer.

\subsection{\label{ssec:gmbom}Generation of Bases for Other Monomers}

The procedure given in section \ref{ssec:gmb} can be used for generating
a contracted basis for each of the  monomers. However, this will not be 
the optimal choice. Because, the molecular symmetry group of the 
cluster includes the subgroup which includes the operations permuting 
the identical monomers as a subgroup. Therefore, after the monomer bases
are combined together they are going to be symmetry adapted by using the 
symmetry operations contained in this group. 
The permutation operations in this group 
will mix the monomer bases, such that when these permutation operations 
are applied to the basis functions of a monomer, then the
resulting function will be a basis function for another monomer. 
If this resulting basis function is not 
already available in the contracted basis of that monomer, then it will
not be orthogonal to the basis functions of that monomer necessarily. 
Therefore, unless there is a relation between the bases of different monomers,
symmetry adaptation to the group which includes the operations 
permuting the identical monomers will always be a 
problem.  As a result of that, 
 a better way of constructing bases for all of the monomers
is to find a basis for one of them,
 and then to generate bases for other monomers
from the basis of this monomer.

The obvious choice for generating the bases for other monomers could be just to 
relabel the basis functions of a single monomer for different monomers. 
However, this will not help  
to get rid of the symmetry adaptation problem posed before. If the  
monomer basis functions are generated with relabeling, 
the basis functions obtained 
in this way will be symmetry adapted in their own permutation groups since
these groups are isomorphic to each other. 
However, if the permutation operations which permute  the
identical  monomers
has some effect on the coordinates other than relabeling them, the action
of these permutation operations to basis functions 
will still result in new basis functions. 
Although relabeling the basis functions of the first monomer to generate
bases for the other monomers can be a solution
for some specific cases, where the action of the permutation operations is 
just to relabel the coordinates (in fact this would be the case if Cartesian
coordinates were used), it is not applicable in general, and is  
far from being a general solution. 

The general solution to that symmetry adaptation problem 
can be found as follows. Firstly, let's consider the case of a dimer.
If the monomers are labeled as $a$ and $b$, then the group which 
contains the permutations of identical monomers will be 
$G_2 =\{E,P_{ab}\}$, where the operation $P_{ab}$ is the 
permutation operation which permutes the monomers $a$ and $b$.
 If $\phi_k^{(a)}$  is the $k^{\mathrm{th}}$ basis 
functions in the optimized basis of the monomer $a$
 and $\phi_l^{(b)}$ is the $l^{\mathrm{th}}$
function in the optimized basis
of the monomer $b$ (which is to be determined),
 then $\phi_k^{(a)} \phi_l^{(b)}$ will 
be one of the product basis functions. In order to make symmetry adaptation
of the basis functions, it will be necessary to apply the permutation operation
$P_{ab}$ to the basis functions. The application of the operation $P_{ab}$
will relabel the basis functions, but it will also introduce some changes
to them so that the new basis function obtained by application 
of this operation will be  
\begin{equation}
 P_{ab}\phi^{(a)}_{k} \phi^{(b)}_{l} = \phi^{(b)'}_{k} \phi^{(a)'}_{l}, 
\label{apppab1}
\end{equation}
where $\phi^{(b)'}_{k}=P_{ab}\phi^{(a)}_{k}$ and 
$\phi^{(a)'}_{l}=P_{ab}\phi^{(b)}_{l}$. The resulting 
basis functions are labeled with $'$s to imply that they are not 
necessarily  in the bases of monomers $a$ and $b$. 
At this point it can be realized that the problem  can be get rid of by 
defining 
\begin{equation}
 \phi^{(b)}_{k} =P_{ab} \phi^{(a)}_{k}.
\label{eq:genbasb}
\end{equation}  
Thus, equation (\ref{apppab1}) becomes,

\begin{eqnarray}
   P_{ab}\phi^{(a)}_k \phi^{(b)}_l & = & P_{ab}( \phi^{(a)}_k (P_{ab} \phi^{(a)}_l))  \\
   & = & (P_{ab}\phi^{(a)}_k) (P_{ab}P_{ab}\phi^{(a)}_l)  \\
   & = & \phi^{(b)}_k \phi^{(a)}_l .
\end{eqnarray}
In the equations above  $P_{ab}P_{ab}=E$ is used that is permuting the 
two monomers between them twice is equivalent to the application of the identity
operation, so that it leaves the system invariant 
(In other words, $P_{ab}$ is its own 
inverse: $P_{ab}^{\dagger}=P_{ab}$). 

Thus, if the basis functions for the monomer $b$ are generated by using 
equation (\ref{eq:genbasb}), then the application of the permutation operation 
$P_{ab}$ does not generate  new basis functions but just carries a
basis function in the basis of a monomer to another basis function in
the basis of the other monomer. 

Although, the discussion above is made for just two monomers, 
the idea can be extended to any bigger cluster.
In the case of a trimer, for example, if the monomers are labeled as 
$a$, $b$ and $c$, the cyclic group 
containing the permutations of identical monomers will be the group
$G_3 = \{E, P_{abc}, P_{acb}\}$.  
In this case the basis of the monomer $b$ can be generated by 
\begin{equation}
   \phi_k^{(b)} = P_{abc} \phi_k^{(a)}.
\end{equation}
The basis of the monomer $c$ can be generated from the basis of monomer
$b$ by 
\begin{equation}
   \phi_k^{(c)} =P_{abc} \phi_k^{(b)}.
\end{equation}
Thus, by repeated application of the generator of the group $G_3$, it is 
possible to generate bases for all of the three monomers
 from the basis of a single
monomer.  
This method can be extended to any bigger cluster. 

If the group $G_l$ 
is the cyclic group with the order $n_m$, then a basis for all of the
monomers can be generated from the basis of a single monomer by 
repeated application of the generator of the group $G_l$.
The reason for the assumption that the group $G_l$ is cyclic should be clear 
at this point. For example, in the case of a trimer if the group 
containing the permutations of the identical monomers would be the group
$G_6 = \{E, P_{abc}, P_{acb}, P_{ab}, P_{ac}, P_{bc}\}$, there would
be more than one way of generating bases for other monomers. For example,
both of the operations $P_{abc}$ and $P_{ab}$ can be used to generate a 
basis for the monomer $b$ from the basis of the monomer $a$. If  
the operation $P_{abc}$ ($P_{ab}$) is used; then, the 
application of the operation
$P_{ab}$ ($P_{abc}$) may still create new basis functions.  
In such a situation, it is impossible to guarantee the invariance of the basis.

Before closing this section, it should also be noted that a basis, which is 
generated by using the generator of the group which includes the operations
that permute identical monomers, will have the same
orthogonality relations with the original basis.
 For example, if one has an orthonormal basis 
for the monomer $a$, then the basis of the monomer $b$ generated by using
equation (\ref{eq:genbasb}) will have 
\begin{eqnarray}
   \langle \phi^{(b)}_k \vert \phi^{(b)}_l \rangle & = & \langle P_{ab} 
                     \phi^{(a)}_k \vert 
            P_{ab} \phi^{(a)}_l \rangle \nonumber \\
   & = & \langle \phi^{(a)}_k \vert P_{ab}^{\dagger} P_{ab} \phi^{(a)}_l \rangle \nonumber \\
   & = & \langle \phi^{(a)}_k \vert \phi^{(a)}_l \rangle \nonumber \\
   & = & \delta_{kl}
\end{eqnarray}

Besides, the basis functions of the monomer $b$ will be eigenstates of the 
model Hamiltonian of the monomer $b$ which is generated in the way that the 
eigenstates of the 
monomer $b$ is generated. Thus, if $\hat{H_a^0}$ is the model 
Hamiltonian of the monomer a, and $\phi^{(a)}_{k}$ is the 
${k}^{th}$ eigenstate of this model Hamiltonian with the eigenvalue 
$\epsilon^{(a)}_{k}$ such that
\begin{equation}
 \hat{H_a^0} \phi^{(a)}_{k} = \epsilon^{(a)}_{k} \phi^{(a)}_{k};
\end{equation}
then,  by applying the permutation operation to both sides of the 
equation above,
\begin{equation}
P_{ab} \hat{H_a^0} \phi^{(a)}_{k}  =  P_{ab}\epsilon^{(a)}_{k}
  \phi^{(a)}_{k}, 
\end{equation}
and inserting the identity operation $P_{ab}^{\dagger}P_{ab}=E$
between Hamiltonian and the basis function
one gets
\begin{equation}
P_{ab}\hat{H_a^0}P_{ab}^{\dagger}P_{ab} \phi^{(a)}_{k} = \epsilon^{(a)}_k 
   P_{ab} \phi^{(a)}_k.
\end{equation}
Thus, 
\begin{equation}
\hat{H_b^0} \phi^{(b)}_{k}  =  \epsilon^{(a)}_{k} \phi^{(b)}_{k},
\label{eigeqmonm}
\end{equation}
where equation (\ref{eq:genbasb}) is used , and $\hat{H_b^0}$ is defined as
\begin{equation}
\hat{H_b^0} = P_{ab} \hat{H_a^0} P_{ab}^{\dagger}.
\end{equation}
Therefore, from equation (\ref{eigeqmonm}), it follows that $\phi^{(b)}_k$
is an eigenstate of the model Hamiltonian $\hat{H_b^0}$ with the eigenvalue
$\epsilon^{(b)}_k = \epsilon^{(a)}_k$, such that 
\begin{equation}
   \hat{H_b^0} \phi^{(b)}_k = \epsilon^{(b)}_k \phi^{(b)}_k.
\end{equation}

\subsection{\label{ssec:cmb}Combining Monomer Bases}

After a basis for one of the monomers is generated by solving for 
the eigenstates of a model Hamiltonian, and the bases for the other
monomers are generated from the basis of this monomer by using the 
generator of the group containing the permutations of identical monomers;
a basis for calculating the spectra of the cluster can be generated by forming
the tensor product of the monomer bases and combining them with a primitive
basis for the inter-monomer coordinates. Thus, if $\phi^{(k)}_{n_k}$ is the
$n^{\mathrm{th}}$ basis function for the monomer $k$ and $\chi_l$ is the 
$l^{\mathrm{th}}$
basis function related with inter-monomer coordinates, the eigenstates of 
the full problem can be expanded in this product basis as  
\begin{equation}
 \Psi_i = \sum_{n_1,n_2,\ldots,n_{n_m},l}C^{n_1, n_2, \ldots,n_{n_m},l}_i 
                 \chi_l \prod_{k=1}^{n_m} \phi^{(k)}_{n_k},
\end{equation}
where $ C^{n_1,n_2,\ldots, n_{n_m},l}_i$ is the expansion coefficient.

As discussed in the previous sections, 
the product basis of the monomer bases will be invariant under the effect of 
any permutation inversion operation. Since a primitive basis is 
used for the inter-monomer coordinates, it can be chosen to be 
invariant under the effect 
of any permutation inversion operation. Consequently, the eigenstates of 
the cluster can be obtained as  linear combinations of the basis functions
of this basis even if the basis is not symmetry adapted for calculations. 
However, symmetry adaptation is always 
useful for reducing the size of the basis. 

The obvious way to make the symmetry adaptation of the basis functions
to an irreducible representation of the molecular symmetry group of a 
cluster is to apply the projection operator of that irreducible representation
to the basis functions. 

Since the group $G_k \otimes \varepsilon$ which is used in monomer 
calculations contains the group $G_k$ as an invariant subgroup, the monomer
basis functions will already be symmetry adapted in their own pure
permutation groups $G_k^{(s)}$. Therefore,
symmetry adaptation of the basis functions can be done in two steps after 
finding the correlations between the irreducible representations of the 
molecular symmetry group and the irreducible representations of the 
pure permutation groups. For example, considering the bases of the monomer $a$
if the irreducible representation 
$\Gamma_{\alpha}$ of the molecular symmetry group correlates to the 
irreducible representation $\Gamma_i$ of the group $G_k^{(a)}$, then 
only the bases which are symmetry adapted to the irreducible representation
$\Gamma_i$ should be used for the calculations. The bases which are
symmetry adapted to the other irreducible representations will be annihilated
by the application of the projection operator of the group $\Gamma_{\alpha}$,
since this projection operator contains the projection operator 
$P^{\Gamma_i}_{G_k^{(a)}}$ of the group $G_k^{(a)}$. On the other hand,
the application of the projection operator $P^{\Gamma_i}_{G_k^{(a)}}$ will
leave the basis functions of the monomer $a$ which are symmetry adapted to the
$\Gamma_i$ irreducible representation of the group $G_k^{(a)}$ invariant. 
Thus, if the correlations between the irreducible representations of the 
molecular symmetry group of the cluster and the pure permutation groups of 
the monomers are found and the appropriate bases are chosen, then the symmetry 
adaptation of the basis functions can be done in two steps by symmetry 
adapting the basis functions further to the groups $G_l$ which is the 
group containing the permutations of the identical monomers and the inversion 
group $\varepsilon$. 

Since monomer basis functions are symmetry adapted to the group 
$G_k \otimes \varepsilon$,  there will be two bases which are
symmetry adapted to two different irreducible representations of the group
$G_k \otimes \varepsilon$ and at the same time 
which are symmetry adapted to the 
irreducible representation $\Gamma_i$ of the group $G_k$. One of these bases
will be symmetry adapted to the irreducible representation 
$\Gamma_i \otimes G=\Gamma_{ig}$, 
which will have even parity, and one of them will be symmetry adapted to the
irreducible representation $\Gamma_i \otimes U =  \Gamma_{iu}$ 
(see table \ref{tab:chartabeps} for the irreducible representations $G$ and
$U$ of the group $\varepsilon$). Thus, if it is necessary to have a basis
for the monomer $a$ which is symmetry adapted to the irreducible representation
$\Gamma_i$, then the basis 
which should be used for the monomer $a$ will be 
\[ \Gamma_{ig} \oplus \Gamma_{iu}, \]
where the labels of the irreducible representations are used to mean any 
basis function belonging to that symmetry. 
The bases of all of the monomers can be found similarly. Thus, when the 
correlations are found and the product basis is formed there will be $2^{n_m}$ 
different product bases
 which differ from each other by the symmetries of monomer 
functions. Since  inter-monomer 
coordinates are usually invariant under the effect of the inversion operation,
half of these terms will have even parity and half of them will have odd
parity.  Thus, when the symmetry adaptation of the basis functions to the
inversion group is done half of these terms will be annihilated and 
half of them will leave invariant. 
After finding the correlations, a fully symmetry adapted
 basis can be generated  by symmetry adapting them  
by using the symmetry operations of the pure permutation group $G_l$. 
Thus,  the symmetry adapted basis functions can be obtained as 
\begin{equation}
   \Psi^{(\Gamma_{\alpha})}= \frac{1}{\vert G_l \vert}\sum_{g \in G_l}
	\chi^{\Gamma_{\alpha}}
         [g]^* \hat{O}_g  \chi_l
      \prod_{k=1}^{n_m} \phi^{(k)}_{n_k},
\end{equation}
where $\Gamma_{\alpha}$ is an irreducible representation of the 
molecular symmetry group,
 and $\chi^{\Gamma_{\alpha}}[g]$
 is the character of the 
operation $g$ represented with the operation $\hat{O}_g$ 
in the irreducible representation $\Gamma_{\alpha}$. 

\subsection{\label{ssec:sol}Solution of The Full Problem}

In order to find the eigenvalues of the Hamiltonian,  the 
matrix elements of the matrix representing the 
Hamiltonian should be calculated.
 They can be evaluated
 easily if the Hamiltonian is partitioned as
\begin{equation}
   \hat{H} = \sum_{k=1}^{n_m} \hat{H_k^0} + \hat{\Delta T } + \hat{\Delta V},
\end{equation}
where $\hat{H_k^0}$'s are the model Hamiltonians for the monomers, 
$\hat{\Delta T}$ is the kinetic energy terms which are not included in 
the model Hamiltonians and 
$\hat{\Delta V}=\hat{V} -\sum_{k=1}^{n_m} \hat{V_k^0}$
is the difference between the potential surface of the full problem and 
the sum of the model potential surfaces used in the model Hamiltonians 
to determine the bases for the monomers. 

The basis functions of the full problem become eigenstates of the zeroth
order Hamiltonian for the full problem, such that by defining
\begin{equation}
   \hat{H^0} = \sum_{k=1}^{n_m} \hat{H_k^0},
\end{equation}
and expanding the wave function in the product basis
\begin{equation}
 \psi_{n_1,n_2\ldots,n_{n_m},l} = \chi_l\prod_{k=1}^{n_m} \phi^{(k)}_{n_k},
\end{equation}
the following eigenvalue relation is obtained 
\begin{equation}
\hat{H^0} \psi_{n_1,n_2,\ldots,n_{n_m},l} = \left(  \sum_{k=1}^{n_m}
                \epsilon^{(k)}_{n_k} \right)
            \psi_{n_1,n_2,\ldots,n_{n_m},l }.
\end{equation}
Therefore, the matrix elements of the Hamiltonian in the basis of 
the full problem will be given by
\begin{eqnarray}
\langle \psi_{n_1',n_2',\ldots,n_{n_m}',l'} \vert \hat{H} \vert
         \psi_{n_1,n_2,\ldots,n_{n_m},l} \rangle = 
         \sum_{k=1}^{n_m} \epsilon^{(k)}_{n_k} \delta_{ll'}
         \prod_{r=1}^{n_m}\delta_{n_rn_r'}  \nonumber \\
         + \langle \psi_{n_1',n_2',\ldots,n_{n_m}',l'} \vert \hat{\Delta T} +
            \hat{\Delta V} \vert \psi_{n_1,n_2,\ldots,n_{n_m},l}\rangle .
\end{eqnarray}
Thus, in order to calculate the matrix elements of the Hamiltonian, 
it is necessary  to evaluate matrix elements of the $\hat{\Delta T}$ and 
$\hat{\Delta V}$ terms,
 in the basis of the full problem. These terms can be evaluated
in the primitive bases of monomers and in the primitive basis of inter-monomer
coordinates, then they can be transformed to the 
contracted basis of the cluster. If the terms  $\hat{\Delta T}$ and 
$\hat{\Delta V}$ are small; then these terms can be considered as a small
perturbation and the basis functions will resemble to the 
eigenstates of the actually problem. In such a case, convergence 
of the results can be 
obtained by using a small number of contracted basis functions. However, this
may not be the case for many problems.

\section{\label{sec:det}Further Details About The MBR Method}

\subsection{\label{ssec:diffmon}The Case of Non-identical Monomers}
While discussing the method in section \ref{sec:mbr}, it was assumed that
the cluster consist of identical monomers. However, this assumption can 
be relaxed easily. 

If there are non-identical monomers in the cluster, then a monomer basis can
be generated for each type of monomer separately in the way it is 
discussed in section \ref{ssec:gmb}. In such a situation, for 
each type of monomer,  there will
be subgroups of the molecular symmetry group  that contain 
permutation operations which permute identical monomers.
The group $G_l$ which is the group that contains the permutations of identical
monomers  will be the direct product of these subgroups. 

After a basis for a single monomer of one type of monomers is 
generated, the procedure of section \ref{ssec:gmbom} can be used to 
generate bases for other monomers which are identical with that monomer.
The same thing can be done for each type of monomers, so that a basis
for each monomer in the cluster is generated. 

Once a basis for each monomer is generated, the procedures of sections
\ref{ssec:cmb} and \ref{ssec:sol} can be used to combine the bases
and to solve the problem. These steps has nothing to do with whether the
monomers are identical or not.  

\subsection{\label{ssec:psspec}Use of The Method with Pseudospectral Methods}

Pseudospectral methods are used frequently in quantum mechanical 
calculations. In this type of methods, two different bases, which are
usually isomorphic to each other, are used for the same degrees of freedom. 
Then, the different parts of the Hamiltonian are evaluated in 
the basis whichever is convenient for that part of the Hamiltonian. 

If pseudospectral methods are used for evaluating the matrix elements of
the Hamiltonian operator, then it becomes necessary to make transformations
from one basis to another. This is achieved by using 
transformation matrices. 
For the monomer $a$, if $\phi_i^{(a)}$ with $i=1,2,\ldots,N$ are 
basis functions in one of the bases and $\theta_i^{(a)}$ with $i=1,2,\ldots,N$ 
are the basis functions in the second basis, then the relation between
the basis functions of the two different bases will be 
\begin{equation}
   \phi_i^{(a)} = \sum_{j=1}^{j=N} T_{ij} \theta_j^{(a)}
\label{eq:ntrans}
\end{equation} 
where $T_{ij}$ are the matrix elements of the transformation matrix $T$.

In section \ref{ssec:gmbom}, it was seen that the proper way of 
generating  bases for other monomers is to use the generator of the group 
which contains the operations that permute identical monomers.
 If the permutation operation $P_{ab}$ is applied to the both
sides of the equation above, by using equation (\ref{eq:genbasb})
 and the fact that
matrix elements are just constants, the equation becomes
\begin{equation}
   \phi_i^{(b)} = \sum_{j=1}^{j=N} T_{ij} P_{ab} \theta_j^{(a)}
\label{eq:phitheta}
\end{equation}
Since the basis functions $\phi_i^{(b)}$ and $\phi_i^{(a)}$ are related to 
each other by equation (\ref{eq:genbasb}) consistency requires that the same
thing should be true for the basis functions $\theta_i^{(b)}$ and 
$\theta_i^{(a)}$, that is 
\begin{equation}
   \theta_i^{(b)} = P_{ab} \theta_i^{(a)}.
\end{equation}
Thus, equation (\ref{eq:phitheta}) becomes
\begin{equation}
   \phi_i^{(b)} = \sum_{j=1}^{j=N} T_{ij} \theta_j^{(b)}.
\label{eq:mtrans}
\end{equation}
Equations (\ref{eq:ntrans}) and (\ref{eq:mtrans})
show that if the basis functions of the monomers are related to each other 
with equation (\ref{eq:genbasb}); 
then, the matrix elements of the transformation
matrix for the bases of the monomer $b$ will be
the same with the matrix elements
of the transformation matrix for the bases of the monomer $a$, so that both
of the 
transformations can be done with the same transformation matrix. Consequently, 
while doing computations, it is not necessary to store separate transformation 
matrices for each monomer. The same transformation matrix can be used for
all of the monomers of the same type.
 Since the transformation matrices usually occupy 
large memories especially if multidimensional coupled bases are
used, this fact is very useful for reducing the memory cost of 
computations. 
 
\section{\label{sec:quamod}Qualitative Models}
In the following two subsections, two qualitative models will be given
for the possible applications of the MBR method. The main discussion will be 
about the symmetries of wave functions. An implementation of the
qualitative model given for water dimer will be made in a following paper
for calculating its vibration-rotation-tunneling spectra.

\subsection{\label{ssec:appwd}Water Dimer}
A group theoretical treatment of water dimer is done by Dyke and co-workers
for explaining the microwave data \cite{dyke77}. The molecular symmetry 
group 
of this dimer is the group $G_{16}$  
which is isomorphic to the 
$D_{4h}$ point group.  If the oxygen atoms in the molecule are labeled as 
$a$ and $b$; the hydrogen atoms
 bonded to the oxygen $a$ are labeled as $1$ and $2$; and the 
hydrogen atoms bonded to the oxygen $b$ are labeled as 
$3$ and $4$; then, 
this group can be written as \cite{MEOPhd} 
\begin{equation}
   G_{16} = \left( \left( G_2^{(a)} \otimes G_2^{(b)} \right) \circledS 
               G_2^{(ab)}\right)
            \otimes \varepsilon ,
\end{equation}
where the monomer permutation groups are 
\begin{eqnarray}
   G_2^{(a)} & = & \{E, (12) \}, \\
   G_2^{(b)} & = & \{E, (34) \},
\end{eqnarray}
the group containing the operations that permute the identical monomers is
\begin{equation}
   G_2^{(ab)} = \{E, (ab)(13)(24) \},
\end{equation}
and the inversion group is 
\begin{equation}
   \varepsilon = \{E, E^* \}.
\end{equation}
 Character table of the group $G_2^{(a)}$ is given in  table 
\ref{tab:chartabg2}.  Character table of the group $G_2^{(b)}$
can be obtained from the character table of the group 
$G_2^{(a)}$ by replacing the operation $(12)$ with the operation $(34)$, 
since the two groups are isomorphic to each other.
Similarly, the character table of the group $G_2^{(ab)}$ can be obtained
from the character table of the group $G_2^{(a)}$ by replacing the operation
$(12)$ with the operation $(ab)(13)(24)$. Character table of the 
group $\varepsilon$ is given in table \ref{tab:chartabeps}.

\begin{table}[p!tb]
\begin{center}
\caption{\label{tab:chartabg2}Character table of the permutation group $G_2$.}
\begin{tabular}{crr}
\hline \hline
$G_2$ & $E$ & $(12)$ \\
\hline
$A$ & $1$ & $1$ \\
$B$ & $1$ & $-1$ \\
\hline \hline
\end{tabular}
\end{center}
\end{table}

\begin{table}[p!tb]
\caption{\label{tab:chartabeps}Character table of the
   inversion group $\varepsilon$.}
\begin{center}
\begin{tabular}{crr}
\hline \hline
$\varepsilon$ & $E$ & $E^*$ \\
\hline
$G$ & $1$ & $1$ \\
$U$ & $1$ & $-1$ \\
\hline
\end{tabular}
\end{center}
\end{table}

\begin{table}[p!tb]
\caption[Character table of the $C_{2v}(M)$ permutation inversion group.]
{\label{tab:chartabc2vm} Character table of the $C_{2v}(M)$
permutation inversion group. This group is the direct product of 
the groups $G_2$ whose character table is given in
table \ref{tab:chartabg2} 
and the group $\varepsilon$ whose character table is given in table
 \ref{tab:chartabeps}. In the table, $\Gamma = x \otimes y$ means that
the irreducible representation $\Gamma$ of the group $C_{2v}(M)$ is obtained
as direct product of the irreducible representation $x$ of the group $G_2$ and
the irreducible representation $y$ of the group $\varepsilon$.}
\begin{center}
\begin{tabular}{crrrr}
\hline \hline 
$C_{2v}(M)=G_2 \otimes \varepsilon$ & $E$ & $(12)$& $E^*$ & $(12)^*$ \\
\hline 
$A_1=A\otimes G$ & 1 & 1 & 1 & 1 \\
$A_2=A\otimes U$ & 1 & 1 & -1 & -1 \\
$B_1=B\otimes U$ & 1 & -1 & -1 & 1 \\
$B_2=B\otimes G$ & 1 & -1 & 1 &  -1 \\
\hline \hline
\end{tabular}
\end{center}
\end{table}

The monomer calculations should be done by symmetry adapting the basis 
functions to the irreducible representations of the  group which is obtained
as a direct product of the pure permutation group of the monomer and the
inversion subgroup of the {\it cluster}. In the case of water dimer, if 
the monomer is chosen as the 
monomer $a$, this means that the monomer calculations
should be done by symmetry adapting the basis functions to the irreducible 
representations of the group  
\begin{equation}
C_{2v}(M)=G_2^{(a)} \otimes \varepsilon = \{E, (12), E^*, (12)^* \}.
\end{equation}
Character table of the group $C_{2v}(M)$ is given in table 
\ref{tab:chartabc2vm}.

After the monomer calculations are done with the basis which is
symmetry adapted to 
the irreducible representations of the group $C_{2v}(M)$, the basis of the
monomer $b$ can be generated by using the generator of the group $G_2^{(ab)}$
which is  $(ab)(13)(24)$. Then,
the question becomes how to combine these bases for the solution of the 
full problem. Character table of the group $G_{16}$, which is the 
molecular symmetry group of water dimer, is given in table 
\ref{tab:chartabg16}. The correlations between the irreducible 
representations of the group $G_{16}$ and the irreducible representations of 
its subgroups are given in table \ref{tab:cortabg16sub}.
 
The monomer bases will already be symmetry adapted to the irreducible 
representations of the groups $G_2^{(a)}$ and $G_2^{(b)}$.
Therefore, 
the product basis of monomer bases will be symmetry adapted in the group
$G_4=G_2^{(a)} \otimes G_2^{(b)}$, whose character table is given in 
table \ref{tab:chartabg4}.

\begin{table}[!tbh]
\caption[Character table of the group $G_4$ which is the subgroup of the
molecular symmetry group of water dimer that contains the operations
permuting identical nuclei within the monomers.]
{\label{tab:chartabg4}  Character table of the group $G_4$.
In the table, $\Gamma=x \otimes y$ means that the irreducible representation
$\Gamma$ of the group $G_4$ is obtained as a direct product of the
irreducible representation $x$ of the group $G_2^{(a)}$ and the irreducible
representation $y$ of the group $G_2^{(b)}$. }
\begin{center}
\begin{tabular}{crrrr}
\hline \hline
$G_4=G_2^{(a)} \otimes G_2^{(b)}$
             & $E$ & $(12)$ & $(34)$ & $(12)(34)$  \\ \hline
$\Gamma_1=A \otimes A $ & $1$  & $1$ & $1$ & $1$ \\
$\Gamma_2=A \otimes B$ & $1$ & $1$ & $-1$ & $-1$  \\
$\Gamma_3=B \otimes A$ & $1$ & -$1$ & $1$ & $-1$  \\
$\Gamma_4=B \otimes B$ & $1$ & -$1$ & $-1$ & $1$  \\
\hline \hline
\end{tabular}
\end{center}
\end{table}


\begin{table*}
\caption[ Character table of
the molecular symmetry group of water dimer.]
{\label{tab:chartabg16}Character table of the $G_{16}$ PI group which is the 
molecular symmetry group of water dimer. This group is isomorphic
to the $D_{4h}$ point group. 
This character table is taken from
a paper by Dyke \cite{dyke77}.}
\begin{center}
\begin{tabular}{crrrrrrrrrr}
\hline \hline
 & & $(12)$ & $(ab)(13)(24)$ & $(ab)(1324)$ & & & $(12)^*$ & $(ab)(13)(24)^*$ & $(ab)(1324)^*$ & \\
$G_{16} $ & $E$ & $(34)$ & $(ab)(14)(23)$& $(ab)(1423)$& $(12)(34)$ & $E^*$ &
            $(34)^*$ & $(ab)(14)(23)^*$ & $(ab)(1423)^*$ & $(12)(34)^*$ \\
\hline
$A_1^+$ & $1$ & $1$ & $1$ & $1$ & $1$ & $1$ & $1$ & $1$ & $1$ & $1$  \\
$A_2^+$ & $1$ & $-1$ & $-1$ & $1$ & $1$ & $1$ & $-1$ & $-1$ & $1$ & $1$  \\
$B_1^+$ & $1$ & $1$ & $-1$ & $-1$ & $1$ & $1$ & $1$ & $-1$ & $-1$ & $1$  \\
$B_2^+$ & $1$ & $-1$ & $1$ & $-1$ & $1$ & $1$ & $-1$ & $1$ & $-1$ & $1$  \\
$E^+$   & $2$ & $0$ & $0$ & $0$ & $-2$ & $2$ & $0$ & $0$ & $0$ & $-2$  \\
$A_1^-$ & $1$ & $1$ & $1$ & $1$ & $1$ & $-1$ & $-1$ & $-1$ & $-1$ & $-1$  \\
$A_2^-$ & $1$ & $-1$ & $-1$ & $1$ & $1$ & $-1$ & $1$ & $1$ & $-1$ & $-1$  \\
$B_1^-$ & $1$ & $1$ & $-1$ & $-1$ & $1$ & $-1$ & $-1$ & $1$ & $1$ & $-1$  \\
$B_2^-$ & $1$ & $-1$ & $1$ & $-1$ & $1$ & $-1$ & $1$ & $-1$ & $1$ & $-1$  \\
$E^-$   & $2$ & $0$ & $0$ & $0$ & $-2$ & $-2$ & $0$ & $0$ & $0$ & $2$  \\
\hline \hline
\end{tabular}
\end{center}
\end{table*}

\begin{table}
\caption[Correlations between the irreducible representations of 
the molecular symmetry group of water dimer $(G_{16})$ and its subgroups.]
{\label{tab:cortabg16sub}Correlations between the irreducible 
representations of the group $G_{16}$ and its subgroups.}
\begin{center}
\begin{tabular}{cccc}
\hline \hline 
$G_{16}$ & $G_4$ & $G_2^{(ab)}$ & $\varepsilon$ \\ \hline
$A_1^+$  & $\Gamma_1$ & $A$ & $G$ \\
$A_2^+$  & $\Gamma_4$ & $B$ & $G$ \\
$B_1^+$  & $\Gamma_1$ & $B$ & $G$ \\
$B_2^+$  & $\Gamma_4$ & $A$ & $G$ \\
$  E^+$  & $\Gamma_2 \oplus \Gamma_3$ & $A \oplus B$ & $G$ \\
$A_1^-$  & $\Gamma_1$ & $A$ & $U$ \\
$A_2^-$  & $\Gamma_4$ & $B$ & $U$ \\
$B_1^-$  & $\Gamma_1$ & $B$ & $U$ \\
$B_2^-$  & $\Gamma_4$ & $A$ & $U$ \\
$  E^-$  & $\Gamma_2 \oplus \Gamma_3$ & $A \oplus B$ & $U$ \\
\hline \hline
\end{tabular}
\end{center}
\end{table}   
 
Considering the $A_1^+$ representation of the group $G_{16}$, 
the correlation table given in table \ref{tab:cortabg16sub} says 
that this level correlates to $\Gamma_1$ irreducible representation
of the group $G_4$, $A$ irreducible 
representation of the group $G_2^{(ab)}$ and 
$G$ irreducible representation of 
the group $\varepsilon$. Therefore, if there is 
a basis for the monomers such that its basis functions are already
 symmetry adapted to the $\Gamma_1$ representation of the group $G_4$, 
then these basis functions can be symmetry adapted to the irreducible 
representations of the group $G_{16}$ by applying the projection 
operators of the $A$ representation of the group $G_2^{(ab)}$ and the 
$G$ representation of the group $\varepsilon$. 

The $\Gamma_1$ irreducible representation of the group $G_4$ is the 
direct product of the $A$ representation of the group $G_2^{(a)}$ and 
the $A$ representation of the group $G_2^{(b)}$. Therefore, the direct 
product of the monomer bases which are symmetry adapted to the $A$ 
representations in their own permutation groups will be symmetry adapted 
to the $\Gamma_1$ irreducible representation of the group $G_4$.

The monomer calculations are done with the group $C_{2v}(M)$. In this 
group both of the irreducible representations $A_1$ and $A_2$ contains 
the $A$ representation of the group $G_2^{(a)}$, therefore both of the 
bases which are symmetry adapted to the irreducible representations 
$A_1$ and $A_2$ of the group $C_{2v}(M)$ can be used for  the monomer 
$a$, so that the basis of the monomer $a$ should be $A_1 \oplus A_2$, where the 
labels of the irreducible representations are used to imply basis functions
belonging to that symmetry.
 The same thing will be true also for the monomer $b$. 
Thus, there will be four different product bases which will be symmetry 
adapted to the irreducible 
representation $\Gamma_1$ of the group $G_4$ which are:
$A_1 \otimes A_1$, $A_1\otimes A_2$, $A_2 \otimes A_1$, $A_2 \otimes A_2$.

The character of the operation $E^*$ is $1$ in the irreducible representation
$A_1$ and $-1$ in the irreducible representation $A_2$. Therefore, the 
basis functions included in the product bases $A_1\otimes A_1$ and 
$A_2 \otimes A_2$ will not change sign upon the operation of the 
inversion operation $E^*$, so that they will have even parity.
On the other hand the basis functions  included
in the product bases $A_1 \otimes A_2$ and $A_2 \otimes A_1$ will change 
sign upon the operation of the inversion operation $E^*$, so that they will
have odd parity.  
In order to fully symmetry adapt the these basis functions to the $A_1^+$
irreducible representation of the group $G_{16}$, it is necessary to apply 
the projection operator of the $G$ irreducible representation of the 
group $\varepsilon$, which is $P^G_{\varepsilon}= (E+E^*)/2$. Since the 
basis functions in the product bases $A_1 \otimes A_1$ and $A_2 \otimes A_2$
does not change sign  upon the operation of the inversion operator, these 
basis functions will be invariant under the effect of the projection
operator $P^G_{\varepsilon}$. On the other hand, since the basis functions
in the product bases $A_1 \otimes A_2 $ and $A_2 \otimes A_1$ change sign 
upon the operation of the inversion operation, they will be annihilated 
when the projection operator $P^G_{\varepsilon}$ is applied. 
Therefore, among the four product bases which are symmetry adapted to 
the $\Gamma_1$ irreducible representation of the group $\Gamma_1$, 
only the bases $A_1 \otimes A_1$ and $A_2 \otimes A_2$ can be used 
for the calculations of the $A_1^+$ representation of the group $G_{16}$.

Thus, the way to generate a basis for the calculations of the $A_1^+$
representations of the group $G_{16}$ is to take the product bases 
$A_1 \otimes A_1$ and $A_2 \otimes A_2$ and symmetry adapt them to the 
irreducible representation $A_1^+$ of the group $G_{16}$ by applying 
the projection operator $P^A_{G_2^{(ab)}}$ of the group $G_2^{(ab)}$. 
Application of the projection operator $P^G_{\varepsilon}$ will 
already leave the basis functions invariant.  

Similarly, the way to combine the monomer bases for the other irreducible
representations of the group $G_{16}$  can be found. The results are
summarized in table \ref{tab:bastabg16}.

\begin{table}
\caption[Symmetries of monomer bases for an MBR calculation of water dimer.]
{\label{tab:bastabg16}This table shows which monomer bases should
be combined for obtaining bases for water dimer calculations with
the group $G_{16}$. In the table, labels of the irreducible representations
are used to imply basis functions belonging to that symmetry. For an explanation
of how to obtain mutually orthogonal basis for the doubly degenerate levels
(i.e. $E^+_x$, $E^+_y$) see reference [\onlinecite{MEOPhd}].}
\begin{center}
\begin{tabular}{cccc}
\hline \hline
$G_{16}$ & Bases & $G_{16}$ & Bases \\ \hline
$A_1^+$  & $(A_1 \otimes A_1) \oplus (A_2 \times A_2)$ &
             $A_1^-$ & $(A_1 \otimes A_2) \oplus (A_2 \otimes A_1)$ \\  
$A_2^+$  & $(B_1 \otimes B_1) \oplus (B_2 \otimes B_2)$ &
             $A_2^-$ & $(B_1 \otimes B_2) \oplus (B_2 \otimes B_1)$ \\  
$B_1^+$  & $(A_1 \otimes A_1) \oplus (A_2 \otimes A_2)$ &
             $B_1^-$ & $(A_1 \otimes A_2) \oplus (A_2 \otimes A_1)$ \\  
$B_2^+$  & $(B_1 \otimes B_1) \oplus (B_2 \otimes B_2)$ &
             $B_2^-$ & $(B_1 \otimes B_2) \oplus (B_2 \otimes B_1)$ \\  
$E^+_x$ & $(A_1 \otimes B_2) \oplus (A_2 \otimes B_1)$ &
             $E^-_x$ & $(A_1 \otimes B_1) \oplus (A_2 \otimes B_2)$ \\  
$E^+_y$ & $(B_2 \otimes A_1) \oplus (B_1 \otimes A_2)$ &
             $E^-_y$ & $(B_1 \otimes A_1) \oplus (B_2 \otimes A_2)$ \\  
\hline \hline 
\end{tabular}
\end{center}
\end{table}

\subsection{\label{ssec:appwt}Water Trimer}
A group theoretical treatment of water trimer is done by van der Avoird {\it et. al.}
\cite{wormer96}. The molecular symmetry group 
of water trimer is the group 
$G_{48}$. 
This group is the direct product of the inversion subgroup 
and the pure permutation subgroup $G_{24}$, such that 
$G_{48}=G_{24} \otimes \varepsilon$. Character table of the group $G_{24}$
is given in table \ref{tab:chartabg24}. In the table $a,b,c$ are the
labels for the oxygen atoms. The hydrogen atoms bonded to 
oxygen $a$ are labeled as 
$1$ and $2$, the hydrogen atoms bonded to oxygen $b$ are labeled as $3$ and $4$,
and the hydrogen atoms bonded to oxygen $c$ are labeled as $5$ and $6$.

\begin{table*}
\caption[{ Character} table of the group $G_{24}$ which is the pure  permutation
subgroup of the molecular symmetry group of water trimer.]
{\label{tab:chartabg24} { Character} table of the group $G_{24}$.
 This table is taken from a paper by van der Avoird {\it et. al.}
 \cite{wormer96}. In the table,
representations given as $A_{2g}$, $A_{3g}$, $A_{2u}$ and $A_{3u}$ in
that article are combined 
to the doubly degenerate representations $E_g = A_{2g} \oplus A_{3g}$ and
$E_u = A_{2u} \oplus A_{3u}$. In contrast to the reference
\cite{wormer96}, operations in the classes are shown explicitly.}
\begin{center}
\begin{tabular}{crrrrrrrr}
\hline \hline
 & & $ (acb)(164253)$ & $(abc)(135264)$ & $(acb)(153)(264)$ & $(abc)(135)(246)$ &  & & \\
 & & $(acb)(153264)$ &  $(abc)(146235)$ & $(acb)(164)(253)$ & $(abc)(146)(235)$
 & $(12)$ & $(12)(34)$ &  \\
 & & $(acb)(154263)$ & $(abc)(145236)$ & $(acb)(154)(263)$ & $(abc)(145)(236)$
    & $(34)$ & $ (12)(56)$ &  \\
 & $E$ & $ (acb)(163254)$ & $ (abc)(136245)$ & $(acb)(163)(254)$
   & $(abc)(136)(245)$ & $(56)$ & $(34)(56)$ & $(12)(34)(56)$ \\
       \hline
$A_{1g}$ & $1$ & $ 1$  & $1$ & $1$ & $1$ & $1$ & $1$ & $1$ \\
$E_g$ & $2$ & $-1$ & $-1$ & $-1$ & $-1$ & $2$  & $2$ & $2$ \\
$T_g$ & $3$ & $0$ & $0$ & $0$ & $0$ & $-1$ & $-1$ & $3$ \\
$A_{1u}$ & $1$ & $-1$ & $-1$ & $1$ & $1$ & $-1$ & $1$ & $-1$ \\
$E_u$ & $2$ & $1$ & $1$ & $-1$ & $-1$ & $-2$ & $2$ & $-2$ \\
$T_u$ & $3$ & $0$ & $0$ & $0$ & $0$ & $1$ & $-1$ & $-3$ \\
\hline \hline
\end{tabular}
\end{center}
\end{table*}

If follows from equation (\ref{eq:gms}), the molecular symmetry group of water trimer can be written in terms of
its subgroups as \cite{MEOPhd} 
\begin{equation}
   G_{48}= \left( \left( G_2^{(a)} \otimes G_2^{(b)} \otimes G_2^{(c)}
         \right) \circledS G_3\right) \otimes \varepsilon,
\end{equation}
where the pure permutation groups of the monomers are
\begin{eqnarray}
   G_2^{(a)} & = & \{ E, (12) \}, \\ 
   G_2^{(b)} & = & \{ E, (34) \} ,\\
   G_2^{(c)} & = & \{ E, (56) \}, 
\end{eqnarray}
and the group containing the operations permuting the monomers is 
\begin{equation}
   G_3 = \{E, (abc)(135)(246), (acb)(153)(264) \}.
\end{equation}
 Character table of the group $G_3$ is given in table \ref{tab:chartabg3}.

\begin{table}[!tbph]
\caption{\label{tab:chartabg3}Character table of the group $G_3$.
 In the table, $w=\exp(2 i \pi/3)$.}
\begin{tabular}{crrr}
\hline \hline
$G_3$ & $E$ & $(abc)(135)(246)$ & $(acb)(153)(264)$ \\ \hline
$A$ & $1$ & $1$ & $1$ \\
$E_x$ & $1$ & $w$ & $w^*$  \\
$E_y$ & $1$ & $w^*$ & $w$ \\
\hline \hline
\end{tabular}
\end{table}

As long as the symmetry adaptation of the monomer calculations are concerned,
there is no difference between water dimer and water trimer. The monomer
calculations should be done by symmetry adapting the basis functions to the
irreducible representations of the group 
$C_{2v}(M)=G_2^{(a)}\otimes \varepsilon$. After the monomer calculations are
done, bases of monomer $b$ can be generated from the bases of
monomer $a$
by using the permutation operation $(abc)(135)(246)$. The bases of the 
monomer $c$ can be generated either from the  bases of the 
monomer $a$ by using the permutation operation $(acb)(153)(264)$ or from the bases of the monomer $b$
by using the permutation operation $(abc)(135)(246)$. 

Since the monomer bases functions are already symmetry adapted in their 
own permutation groups, the product basis of the monomer bases will be 
symmetry adapted in the group 
$G_8=G_2^{(a)} \otimes G_2^{(b)} \otimes G_2^{(c)}$. Character table of the
group $G_8$ is given in table \ref{tab:chartabg8trimer}.  The correlations
between the irreducible representations of the group $G_{48}$ and its 
subgroups $G_8$, $G_3$ and $\varepsilon$ are given in table
\ref{tab:cortabg48sub}.

\begin{table*}
\caption[{ Character} table of the group $G_8$ which is the subgroup of the
molecular symmetry group of water trimer that contains the operations
permuting the identical nuclei within monomers.]
{\label{tab:chartabg8trimer} Character table of the group $G_8$.
 Correlations with the irreducible representations of the subgroups
are indicated such that in the table
$\Gamma=x \otimes y \otimes z$ means that the
irreducible representation $\Gamma$ of the group $G_8$ is obtained as a
direct product of the irreducible representations $x$, $y$ and $z$  of the
groups $G_2^{(a)}$, $G_2^{(b)}$ and $G_2^{(c)}$, respectively.}
\begin{center}
\begin{tabular}{crrrrrrrr}
\hline \hline
$G_8=G_2^{(a)} \otimes G_2^{(b)} \otimes G_2^{(c)}$ & $E$ & $(12)$
 & $ (34)$ & $ (56)$ & $(12)(34)$ & $(12)(56)$ & $(34)(56)$ & $(12)(34)(56)$ \\ \hline
$\Gamma_1=A \otimes A \otimes A$
               & $1$ & $1$ & $1$ & $1$ & $1$ & $1$ & $1$ & $1$\\
$\Gamma_2=A \otimes A \otimes B$
               & $1$ & $1$ & $1$& $-1$& $1$ & $-1$& $-1$& $-1$\\
$\Gamma_3=A \otimes B \otimes A$
               & $1$ & $1$ & $-1$& $1$ & $-1$& $1$ & $-1$& $-1$\\
$\Gamma_4=B \otimes A \otimes A$
               & $1$ & $-1$ & $1$& $1$ & $-1$ & $-1$ & $1$ & $-1$\\
$\Gamma_5=A \otimes B \otimes B$
               & $1$ & $1$& $-1$ & $-1$ & $-1$& $-1$& $1$ & $1$\\
$\Gamma_6=B \otimes A \otimes B$
               & $1$ & $-1$& $1$ & $-1$& $-1$& $1$ & $-1$ & $1$\\
$\Gamma_7=B \otimes B \otimes A$
               & $1$ & $-1$& $-1$& $1$ & $1$ & $-1$ & $-1$ & $1$\\
$\Gamma_8=B \otimes B \otimes B$
               & $1$ & $-1$ & $-1$ & $-1$ & $1$ & $1$ & $1$ & $-1$\\
\hline \hline
\end{tabular}
\end{center}
\end{table*}

In order to illustrate how the monomer bases can be combined for the 
solution of the full problem consider the $E_g^+$ representation. 
This representation correlates to the $\Gamma_1$ 
representation of the group $G_8$. Since the irreducible representation 
$\Gamma_1$ is obtained as a direct product of the $A$ representations of 
the pure permutation groups of the monomers (see table \ref{tab:cortabg48sub}),
  the product basis of the 
monomer bases which are symmetry adapted to the $A$ representation should
be used for forming the monomer bases for the solution of the full problem.
Since the monomer calculations are done with the group $C_{2v}(M)$, both
of the $A_1$ and $A_2$ representations are symmetry adapted to the $A$
representation of the pure permutation group of the monomers. Therefore, 
the monomer bases that should be used in the calculations is the sum of 
 $A_1$ and $A_2$ bases which is $A_1 \oplus A_2$, and the product basis 
of the monomer bases is going to be 
\[ ( A_1 \oplus A_2) \otimes (A_1 \oplus A_2) \otimes (A_1 \oplus A_2).\]  
When the multiplications are done, there will be eight terms. 
Half of them will have even parity and half of
them will have odd parity. The functions having odd parity will be 
annihilated upon the operation of the projection operator $P^G_{\varepsilon}$
of the inversion group. On the other hand the functions having even parity 
will be invariant under the effect of the projection operator
$P^G_{\varepsilon}$. Thus, only the terms which have the right parity will
lead to basis for the full problem. These bases are 
$(A_1 \otimes A_1 \otimes A_1) \oplus (A_1 \otimes A_2 \otimes A_2) 
 \oplus (A_2 \otimes A_1 \otimes A_2 )  \otimes (A_2 \otimes A_2 \otimes A_1)$.
After these bases are combined with a basis related with inter-monomer 
coordinates, they can be symmetry adapted to the $E_g^+$ irreducible 
representation of the group $G_{48}$ by application of the 
projection operator $P^{E_x}_{G_3} + P^{E_y}_{G_3}$.

Similarly, the monomer bases which should be used to form a product basis
for each irreducible representation of the group $G_{48}$ can be found. 
The results are given in table \ref{tab:bastabg48}.

\begin{table}
\caption[Correlations between the irreducible representations of the 
molecular symmetry group of water trimer $(G_{48})$ and its subgroups.]
{\label{tab:cortabg48sub}Correlations between the irreducible
representations of the group $G_{48}$ and its subgroups.}
\begin{center}
\begin{tabular}{cccc}
\hline \hline 
$G_{48}$ & $G_8$ & $G_3$ & $\varepsilon$ \\ \hline 
$A_{1g}^+$ &  $\Gamma_1$ & $A$ & $G$ \\
$E_g^+$    &  $2*\Gamma_1$ & $E_x \oplus E_y$ & $G$ \\
$T_g^+$    &  $\Gamma_5 \oplus \Gamma_6 \oplus \Gamma_7$ & 
                           $A\oplus E_x \oplus E_y$ & $G$ \\
$A_{1u}^+$ &  $\Gamma_8$ & $A$ & $G$ \\
$E_u^+$    &  $2*\Gamma_8$ & $E_x \oplus E_y$ & $G$ \\
$T_u^+$    &  $\Gamma_2 \oplus \Gamma_3 \oplus \Gamma_4$ & 
                           $A \oplus E_x \oplus E_y$ & $G$ \\
$A_{1g}^-$ &  $\Gamma_1$ & $A$ & $U$ \\
$E_g^-$    &  $2*\Gamma_1$ & $E_x \oplus E_y$ & $U$ \\
$T_g^-$    &  $\Gamma_5 \oplus \Gamma_6 \oplus \Gamma_7$ & 
                           $A\oplus E_x \oplus E_y$ & $U$ \\
$A_{1u}^-$ &  $\Gamma_8$ & $A$ & $U$ \\
$E_u^-$    &  $2*\Gamma_8$ & $E_x \oplus E_y$ & $U$ \\
$T_u^-$    &  $\Gamma_2 \oplus \Gamma_3 \oplus \Gamma_4$ & 
                           $A \oplus E_x \oplus E_y$ & $U$ \\
\hline \hline
\end{tabular}
\end{center}
\end{table}

\begin{table*}[ptbh]
\caption[Symmetries of monomer bases for an MBR calculation of water trimer.]
{\label{tab:bastabg48}This table shows which monomer bases should 
be combined for obtaining bases for the calculations of the water trimer
with the group $G_{48}$. In the table labels of the irreducible representations
are used to imply the monomer basis functions belonging to that symmetry.
For an explanation of how to obtain mutually orthogonal bases for the 
triply degenerate levels (i.e. $T_{gx}^+$, $T_{gy}^+$, $T_{gz}^+$), see 
reference [\onlinecite{MEOPhd}].} 
\begin{tabular}{cr}
\hline \hline 
$G_{48}$ Bases \\ \hline 
$A_{1g}^+$ $(A_1\otimes A_1 \otimes A_1) \oplus (A_1 \otimes A_2 \otimes A_2)
  \oplus (A_2 \otimes A_1 \otimes A_2) \oplus (A_2 \otimes A_2 \otimes A_1)$  \\
$E_{g}^+$ $(A_1\otimes A_1 \otimes A_1) \oplus (A_1 \otimes A_2 \otimes A_2)
  \oplus (A_2 \otimes A_1 \otimes A_2) \oplus (A_2 \otimes A_2 \otimes A_1)$  \\
$T_{gx}^+$ $(A_1\otimes B_2 \otimes B_2) \oplus (A_1 \otimes B_1 \otimes B_1)
  \oplus (A_2 \otimes B_2 \otimes B_1) \oplus (A_2 \otimes B_1 \otimes B_2)$  \\
$T_{gy}^+$ $(B_2\otimes B_2 \otimes A_1) \oplus (B_1 \otimes B_1 \otimes A_1)
  \oplus (B_2 \otimes B_1 \otimes A_2) \oplus (B_1 \otimes B_2 \otimes A_2)$  \\
$T_{gz}^+$ $(B_2\otimes A_1 \otimes B_2) \oplus (B_1 \otimes A_1 \otimes B_1)
  \oplus (B_1 \otimes A_2 \otimes B_2) \oplus (B_2 \otimes A_2 \otimes B_1)$  \\
$A_{1u}^+$ $(B_2\otimes B_2 \otimes B_2) \oplus (B_2 \otimes B_1 \otimes B_1)
  \oplus (B_1 \otimes B_2 \otimes B_1) \oplus (B_1 \otimes B_1 \otimes B_2)$  \\
$E_{u}^+$ $(B_2\otimes B_2 \otimes B_2) \oplus (B_2 \otimes B_1 \otimes B_1)
  \oplus (B_1 \otimes B_2 \otimes B_1) \oplus (B_1 \otimes B_1 \otimes B_2)$  \\
$T_{ux}^+$ $(A_1\otimes A_1 \otimes B_2) \oplus (A_1 \otimes A_2 \otimes B_1)
  \oplus (A_2 \otimes A_1 \otimes B_2) \oplus (A_2 \otimes A_2 \otimes B_2)$  \\
$T_{uy}^+$ $(A_1\otimes B_2 \otimes A_1) \oplus (A_2 \otimes B_1 \otimes A_1)
  \oplus (A_1 \otimes B_2 \otimes A_2) \oplus (A_2 \otimes B_2 \otimes A_2)$  \\
$T_{uz}^+$ $(B_2\otimes A_1 \otimes A_1) \oplus (B_1 \otimes A_1 \otimes A_2)
  \oplus (B_2 \otimes A_2 \otimes A_1) \oplus (B_2 \otimes A_2 \otimes A_2)$  \\
$A_{1g}^-$ $(A_2\otimes A_2 \otimes A_2) \oplus (A_1 \otimes A_1 \otimes A_2)
  \oplus (A_1 \otimes A_2 \otimes A_1) \oplus (A_2 \otimes A_1 \otimes A_1)$  \\
$E_{g}^-$ $(A_2\otimes A_2 \otimes A_2) \oplus (A_1 \otimes A_1 \otimes A_2)
  \oplus (A_1 \otimes A_2 \otimes A_1) \oplus (A_2 \otimes A_1 \otimes A_1)$  \\
$T_{gx}^-$ $(A_2\otimes B_1 \otimes B_1) \oplus (A_1 \otimes B_2 \otimes B_1)
  \oplus (A_1 \otimes B_1 \otimes B_2) \oplus (A_2 \otimes B_2 \otimes B_2)$  \\
$T_{gy}^-$ $(B_1\otimes B_1 \otimes A_2) \oplus (B_2 \otimes B_1 \otimes A_1)
  \oplus (B_1 \otimes B_2 \otimes A_1) \oplus (B_2 \otimes B_2 \otimes A_2)$  \\
$T_{gz}^-$ $(B_1\otimes A_2 \otimes B_1) \oplus (B_1 \otimes A_1 \otimes B_2)
  \oplus (B_2 \otimes A_1 \otimes B_1) \oplus (B_2 \otimes A_2 \otimes B_2)$  \\
$A_{1u}^-$ $(B_1\otimes B_1 \otimes B_1) \oplus (B_2 \otimes B_2 \otimes B_1)
  \oplus (B_2 \otimes B_1 \otimes B_2) \oplus (B_1 \otimes B_2 \otimes B_2)$  \\
$E_{u}^-$ $(B_1\otimes B_1 \otimes B_1) \oplus (B_2 \otimes B_2 \otimes B_1)
  \oplus (B_2 \otimes B_1 \otimes B_2) \oplus (B_1 \otimes B_2 \otimes B_2)$  \\
$T_{ux}^-$ $(A_2\otimes A_2 \otimes B_1) \oplus (A_1 \otimes A_1 \otimes B_1)
  \oplus (A_1 \otimes A_2 \otimes B_2) \oplus (A_2 \otimes A_1 \otimes B_2)$  \\
$T_{uy}^-$ $(A_2\otimes B_1 \otimes A_2) \oplus (A_1 \otimes B_1 \otimes A_1)
  \oplus (A_2 \otimes B_2 \otimes A_1) \oplus (A_1 \otimes B_2 \otimes A_2)$  \\
$T_{uz}^-$ $(B_1\otimes A_2 \otimes A_2) \oplus (B_1 \otimes A_1 \otimes A_1)
  \oplus (B_2 \otimes A_1 \otimes A_2) \oplus (B_2 \otimes A_2 \otimes A_1)$  \\
\hline \hline
\end{tabular}
\end{table*}

\section{Conclusions}

By using the sequential symmetry adaptation procedure, and the physically
meaningful partitioning of the molecular symmetry group of molecular 
clusters, 
a new method for calculating the VRT spectra of molecular clusters
named Monomer Basis Representation (MBR) method was 
developed in section \ref{sec:mbr}. 
In the MBR method, calculations starts with a single monomer with 
the purpose of obtaining an optimized basis for that monomer
as a linear combination of some primitive basis functions. Then, 
an optimized basis for each identical monomer is generated from the
optimized basis of this monomer. By using the optimized bases of 
the monomers, a basis is generated  for the solution of
the full problem, and the VRT spectra of the cluster is obtained 
by using this basis. Since an optimized basis is used for each monomer
which has a much smaller size than the primitive basis from which 
the optimized bases are generated, the MBR method leads to an exponential
decrease in the size of the basis that is required for the 
convergence of the results. 

The generation of an optimized basis for each monomer as a linear combination
of some primitive basis functions led to two problems  related with the 
symmetry adaptation. When a basis for the solution of the full problem 
is obtained as a direct product of the optimized monomer bases. These basis
functions should still be symmetry adapted to the inversion group and also
to the group which contains the permutations of the monomers. 
In order to symmetry adapt basis functions to these groups, it is necessary to
apply the symmetry operations contained in these groups 
to the basis functions.  When these
operations are applied to the basis functions, the resulting functions should
already be in the basis in order to have an orthogonal basis. However, when 
the monomer basis functions are linear combinations of some primitive basis
functions, the functions obtained as a result of the operations of the
permutation inversion operations will not be necessarily in the same basis.

It has been suggested that the invariance of the basis 
functions under the effect of the inversion operation can be achieved 
by symmetry adapting primitive monomer basis functions to the inversion 
subgroup while generating the optimized basis for the monomers. If the 
monomer basis functions are already symmetry adapted to the inversion subgroup,
then they will be eigenstates of the inversion 
operator with the eigenvalues $\pm 1$. Consequently, the basis functions 
which are obtained as a direct product of the monomer basis  functions
will be  the eigenstates of the inversion operator with the eigenvalues $\pm 1$.
Thus, the basis which is obtained as  a direct product of the optimized 
monomer bases becomes invariant under the effect of the inversion operation. 

It has also been shown that the product basis of the optimized monomer 
bases can be made invariant under the effect of the operations permuting 
identical monomers by finding an optimized basis for a single monomer 
and then generating bases for other monomers from the basis of 
that monomer. If an  
optimized basis for each  monomer is generated separately independent of 
other monomer bases, then there is no way to guarantee the invariance of
the basis. On the other hand, if all of the monomer bases are generated from
a single monomer basis, then it is possible to correlate the bases to each 
other so that the product basis of optimized monomer bases becomes invariant.
It has been shown that a basis for all of the monomers can be generated 
from the basis on a single monomer by repeated application of the 
generator of the group $G_l$ which is the group containing the operations
that permute identical monomers. This way of generating bases for all of the
monomers made it necessary to assume that the order of the group $G_l$ is
equal to the number of the monomers: $G_l=n_m$.

While developing the method,
 the primitive bases of the monomers were never referenced. 
Thus, the nature of the primitive monomer bases does not matter. They can 
be product bases of one dimensional bases, or they can be 
multidimensional coupled bases. The symmetry adaptation procedure given 
here will work regardless of the nature of the primitive bases used in 
the monomer calculations. 

Another point which is not discussed in detail in the development of the 
method was what should be the model 
potential surface while generating a monomer basis. This is just 
because it is impossible to suggest a perfect model potential surface regardless
of the particular problem being studied.
 The model potential surface should be chosen 
according to the problem at hand. Nevertheless it should be noted
that the more the model potential resembles to the actual problem, the more
efficient the monomer bases will be. 

\appendix

\section*{Decomposition Of Projection Operators}

In order to find a relation between the projection operators of a 
product group and the projection operators of its subgroups, it is 
necessary to find a relation between the characters of the elements
of the product group and the characters of the elements of the subgroups. 
Such a relation can be derived from the following character equation
which holds in any irreducible representation of any group \cite{TinkCharRel}:
\begin{equation}
N_i \chi(C_i) N_j \chi (C_j) = d \sum_k c_{ijk} N_k \chi (C_k). 
\label{eq:charrel}
\end{equation}
In the equation above, $C_i$, $C_j$ and $C_k$ are classes of the group;
$N_i$, $N_j$ and $N_k$ are the number of elements in these classes; and
the coefficients $c_{ijk}$ are defined by the class multiplication 
equation $C_i C_j = \sum_k c_{ijk} C_k$; and $d$ is the dimension of the
irreducible representation.  

Consider a group $G$ that can be written as a semi-direct product of two of its 
subgroups $H$ and $K$ such that $G=H\circledS K$. If $h$ is an element
of group $H$ that is in class $C_i$ of the group $G$ and $k$ is an element
of the group $K$ that is in class $C_j$ of the group $G$, and
$g=hk$ is an element of the group $G$ that is in class $C_m$;
 then provided that
the class multiplication constants satisfy the equation 
\begin{equation}
c_{ijk} = r \delta_{km}, \label{eq:cijk}
\end{equation}
where $r$ is an integer; 
equation (\ref{eq:charrel}) reduces to
\begin{equation}
\chi(h) \chi(k) = d \chi(g), \label{eq:chdecomp}
\end{equation}
and equation (\ref{eq:pod}) follows.
Consequently, equation (\ref{eq:cijk}) is the sufficient condition that
the sequential symmetry adaptation will work for any irreducible
representation of the product group. This equation seems to be 
satisfied in many cases for the physically meaningful partitioning of the
molecular molecular symmetry group of molecular clusters given in
equation (\ref{eq:gms}). The examples include molecular symmetry 
groups of the clusters $(H_2O)_2$, $(CO_2)_2$, $(H_2O)_3$, $(H_2O)_2 D_2O$.
However, although it has been argued before \cite{MEOPhd} that the equation 
(\ref{eq:cijk}) holds for any irreducible representation of any semi-direct
product group, this is not the case. For example, if pairwise permutations
of monomers would be a feasible symmetry operation for water trimer so that
the molecular symmetry group would be $G_{96}$ instead of $G_{48}$, then
equation (\ref{eq:cijk}) would not hold for the semi-direct product
multiplication defined by equation (\ref{eq:gms}).
 
It should also be noted that equation (\ref{eq:chdecomp}) holds for 
any one dimensional representation. This follows from the fact that
one dimensional representations are representations by nonzero complex
numbers and a representation should satisfy equation (\ref{eq:chdecomp})
by definition of representation since the characters of the elements are 
just the complex numbers representing them. 
Consequently, even if the condition given in equation (\ref{eq:cijk})
does not hold, sequential symmetry adaptation can 
still be used for one dimensional
representations.

\newpage

\bibliography{bibliography}

\end{document}